\documentstyle[preprint,tighten,eqsecnum,aps,floats,psfig,epsfig]{revtex}
\def\ffrac#1#2{\textstyle{#1\over#2}\displaystyle}
\def\perm{{\rm perm}\,}
\newcommand{\qed}{\hbox{\rule{6pt}{10pt}}}

\begin{document}
\title{Network Models in Class C on Arbitrary Graphs}
\author{John Cardy}
\address{Rudolph Peierls Centre for Theoretical Physics\\
         1 Keble Road, Oxford OX1 3NP, U.K.\footnote{Address
for correspondence} \\
         and All Souls College, Oxford.\\}
%
\maketitle
\begin{abstract}
We consider network models of quantum localisation in which a particle
with a two-component wave function propagates through the nodes and
along the edges of an arbitrary directed graph, subject to a random
SU$(2)$ rotation on each edge it traverses. 
The propagation through each node is specified by
an arbitrary but fixed $S$-matrix. Such networks model
localisation problems in class C of the classification of Altland and
Zirnbauer, and, on suitable graphs, they model the spin quantum Hall
transition. We extend the analyses of Gruzberg, Ludwig and Read and of
Beamond, Cardy and Chalker to show that, on an arbitrary graph, the mean
density of states and the mean conductance may be calculated in terms
of observables of a classical history-dependent random walk on the same
graph. The transition weights for this process are explicitly related to
the elements of the $S$-matrices. They are correctly normalised but, on
graphs with nodes of degree greater than 4, not necessarily non-negative
(and therefore interpretable as probabilities) unless a sufficient
number of them happen to vanish. Our methods use a supersymmetric path integral
formulation of the problem which is completely finite and
rigorous. 
\end{abstract}
%
%
\section{Introduction}
Network models of quantum localisation were first introduced by
Chalker and Coddington\cite{CC} to model
the transition between plateaux in integer quantum Hall systems.
Reduced to their essentials, they describe the propagation of a single
quantum-mechanical particle along the directed edges and through the nodes of
a graph. For the Chalker-Coddington model, this graph is some large but
bounded domain of the L-lattice, a square lattice whose edges are
directed in such a way that the particle turns through $\pm 90^{\circ}$
at each node. In propagating along each edge, the single-component
wave function is
multiplied by random phases, which are i.i.d. random variables with a
uniform distribution in $[0,2\pi)$. The propagation through each
node is described by a unitary
$2\times2$ $S$-matrix of amplitudes between the two
incoming and two outgoing edges. 

The integer quantum Hall plateau transition is only one among several
possible universality classes of quantum localisation transitions, which
have been classified by Altland and Zirnbauer\cite{AZ} according to the
symmetry properties of the underlying single-particle 
hamiltonian $\cal H$. Another, known as class C, corresponds to the
existence of a symmetry 
\begin{equation}
\label{symm}
\sigma_y{\cal H}\sigma_y=-{\cal H}^*\quad.
\end{equation}
This gives rise to a pairing between eigenstates with energies $\pm E$,
while $E=0$ is special and may correspond to delocalised eigenstates,
even in two dimensions. Class C is supposed to be realised in
disordered spin-singlet superconductors in which time-reversal symmetry
is broken, but Zeeman splitting is negligible\cite{AZ}. The fact that spin is 
still conserved can then lead to a spin quantum Hall effect. 

The appropriate network model for this on the L-lattice was formulated 
and studied numerically by Kagalovsky et al\cite{Kag}. An equivalent
spin-chain hamiltonian was also investigated by Senthil et
al\cite{Sent}.
However, in a remarkable paper, Gruzberg, Ludwig and Read\cite{GLR} showed that
the mean single-particle Green function, as well as the mean
conductance, may be expressed in terms of \em classical \em averages
of appropriate observables of the hulls (boundaries) of clusters
in classical bond percolation on the square lattice. The critical
exponents\cite{GLR} and some other universal properties\cite{JCcond}
of the spin quantum Hall
transition in two dimensions are thus exactly known. 

The methods of Gruzberg et al\cite{GLR} used supersymmetry to average over
quenched disorder. One of the essential features of this, which will
also appear in our analysis, is the reduction of the Hilbert space on
each edge to one of finite dimension. They then analysed the transfer
matrix for the L-lattice, and demonstrated its equivalence to that for
percolation hulls. 

One of the interesting features of percolation hulls on the square
lattice is that they may be generated, independently of the underlying
percolation problem, as history-dependent random walks on the L-lattice.
Consider a random walk which begins on some edge, and steps through one
node in unit time. At each node it turns to the left or right with
probabilities $p$ or $1-p$ respectively. However, it cannot traverse a
given edge more than once, so that whenever it returns to a node it has
already visited it is forced to exit along the other (empty) edge. 
Eventually, on a closed graph, it will return to its initial edge.
The statistical properties of such loops are identical to those of a
single closed hull in the percolation problem. On the L-lattice in the
thermodynamic limit, when $p\not=\frac12$, loops close almost surely
after a finite number of steps. This corresponds to quantum localisation
in the network model. The only delocalised states occur when
$p=\frac12$, at the bond percolation threshold. 

Motivated by this work, Beamond, Cardy and Chalker\cite{BCC}
investigated class C network models on arbitrary graphs. Their methods
did not use supersymmetry. Instead they showed that, for quantities like
the average Green function $G$ and the mean conductance (related to 
$|G|^2$), there is a massive cancellation between paths in the Feynman
expansion of these quantities which leaves essentially classical paths
whose weights correspond to those of a history-dependent random walk. 
This reproduces the results of Gruzberg et al\cite{GLR} when specialised to the
L-lattice. 

However, the proof of Beamond et al\cite{BCC} was restricted to graphs in which
each node has $N=2$ incoming and outgoing directed edges. 
It becomes too cumbersome to generalise it to graphs with nodes with
$N>2$. 
Nevertheless, it seems important to find such a generalisation in order
to be able to investigate, for example, the properties of such network
models on simple regular lattices embedded in three or more dimensions. 
 
In this paper, we find this generalisation. Like Gruzberg et al\cite{GLR}, we
use supersymmetry to perform the quenched average. However, we do this
within a path integral, rather than a Hilbert space, formulation of the
model, which allows for the treatment of an arbitrary graph, not only
those regular lattices which admit a transfer matrix. 
The result is positive, in the
sense that we can prove that the mean of $G$ and of $|G|^2$ may be expressed
as a sum of history-dependent classical random walks on the graph. The
weights at each node, given that a given set
of incoming edges is occupied by the walk, when summed over all possible
outcomes, correctly sum to unity.
However, in general these weights are not positive, and therefore cannot
be interpreted as probabilities. In fact we show that, for $N>2$, this
condition can only be satisfied if a certain number of elements of $S$
vanish. A sufficient condition for this is that the
$S$-matrix at each node is a direct product of $S$-matrices for $2\to2$
nodes, that is, the node may be decomposed into $2\to2$ nodes (in which
case, of course, the analysis of Beamond et al\cite{BCC} applies.)
For $N=3$ and $4$ we have shown that this condition is also necessary.

Despite this negative result, 
the proof of the general theorem sheds further light on
the analysis of Gruzberg et al\cite{GLR}, as well as giving a more elegant
derivation of Beamond et al\cite{BCC}. 
These methods may also be used to determine which combinations of
averages of higher-point Green functions in the network model may be
related to observables in the classical problem, thus giving a simpler
derivation of some of the results of Mirlin, Evers, and
Mildenberger\cite{Mirlin}.

The layout of this paper is as follows. In the next Section, we define
the network models and the observables of interest. Then we are able to
state our main Theorems (1 and 2). In Sec.~III we introduce the path
integral machinery necessary to compute these observables, and to
perform quenched averages. The supersymmetric path integral involves
both bosonic (commuting) and fermionic (anticommuting) variables on each
edge, but after the quenched average is taken we show that these reduce
to the propagation of only single fermion-fermion or boson-fermion pairs.
In Sec.~IV we then show that the propagation of a fermion-fermion pair
through the lattice obeys the rules of a classical history-dependent
random walk, once all the other degrees of freedom are traced out.
This proves Theorem 1. We also consider the case of open systems, and
conductance measurements. Finally, in Sec.~V, we consider the
probabilistic interpretation in terms of history-dependent random
walks, and prove Theorem 4 which states conditions under which these
weights are non-negative.
\section{Definition of the model and observables.}

Let $\cal G$ be a graph consisting of $\cal N$ directed edges, and nodes.
Initially we consider only closed graphs. At each node there is an equal
number of incoming and outgoing directed edges. Apart from this, $\cal
G$ is arbitrary. We wish to define a network model on this graph with
describes the propagation of a quantum-mechanical particle whose
hamiltonian $\cal H$ obeys the symmetry (\ref{symm}). First we note that
the single-particle Hilbert space must be even-dimensional in order to
be able to define the action of $\sigma_y$. For simplicity we take this
to be two-dimensional (in \cite{BCC} a method was proposed for reducing
any class C network model with an even-dimensional single-particle Hilbert
space to this case.) Since the network describes propagation over finite
time steps $\Delta t$, we need to define the unitary evolution operators
$U_e$ and $U_n$ which evolve the wave function along each edge and node
respectively. By (\ref{symm}), they must obey $\sigma_yU\sigma_y=U^*$. 
Each $U_e$, therefore, must be an element of Sp(2)$\simeq$SU(2).
We take these to be i.i.d. random variables, each uniformly distributed
with respect to the Haar measure of SU(2). Quenched averages with
respect to this measure will be denoted by an overline.
$U_n$ is the unitary $S$-matrix for the node $n$. As will become clear,
it suffices to take this to be diagonal in the SU(2) indices, so,
by the above, it must be real and therefore 
an element of O$(N)$, where $N$ is the number of incoming (and outgoing)
edges at the node. The $U_n$ may vary from node to node, and in principle they 
may also be random variables. 
However, our theorems apply to a fixed realisation of these $S$-matrix
elements.

The full unitary evolution operator $\cal U$ for the
whole network is then a direct sum over the edges and nodes of
${\bf 1}\otimes\ldots\otimes U_e\otimes\ldots\otimes{\bf 1}$
and ${\bf 1}\otimes\ldots\otimes U_n\otimes\ldots\otimes{\bf 1}$.
Of interest is the Green function, which is the matrix element of 
the resolvent operator $(1-z{\cal U})^{-1}$ between states localised on
two edges:
\begin{equation}
\label{G}
G(e_2,e_1;z)\equiv\langle e_2|(1-z{\cal U})^{-1}|e_1\rangle\quad.
\end{equation} 
This is a $2\times2$ matrix in SU(2) space. For $|z|<1$ this may be
expanded as a sum of Feynman paths on $\cal G$ beginning at $e_1$ and ending 
at $e_2$: each path is weighted by an ordered product of SU(2)
matrices $U_e$ along the edges is traverses, and a product of $S$-matrix
elements according to how it passes through each node, as well as a
factor $z$ raised to the power of its length. However, each edge may be
traversed an arbitrary number of times. If we were to formulate the sum
over such paths using a transfer matrix formalism (assuming that $\cal
G$ allows this), the Hilbert space on each edge would be
infinite-dimensional. 

Alternatively, for $|z|>1$, we may write the resolvent as 
$-z^{-1}{\cal U}^{\dag}(1-z^{-1}{\cal U}^{\dag})^{-1}$ and expand $G$
in powers of $z^{-1}$, as a similar sum over paths. Each path is now
weighted by an ordered product of factors $z^{-1}U_e^{\dag}$, as well as
an overall factor of $-z^{-1}$. 

The Green function may be used to compute the density of states of $\cal
U$ via its diagonal elements $G(e,e;z)$. For a closed graph, these are
of the form $\exp(i\epsilon_j)$. We define the density of states
$\rho(\epsilon)=\sum_j\delta(\epsilon-\epsilon_j)$. Then
\begin{equation}
\rho(\epsilon)={1\over2\pi}{1\over2{\cal N}}\sum_e\lim_{\delta\to0}
\left({\rm Tr}\,G(e,e;(1-\delta)e^{-i\epsilon})-
{\rm Tr}\,G(e,e;(1+\delta)e^{-i\epsilon})\right)
\end{equation}

An open system may be defined by cutting open a subset of the edges of
$\cal G$. We may then perform a conductance measurement by attaching
leads to a subset $\{e_{\rm in}\}$ of the incoming edges, and to a
subset $\{e_{\rm out}\}$ of the outgoing edges.
The transmission matrix $t$ between these two leads has elements
$\langle e_{\rm out}|(1-{\cal U})|e_{\rm in}\rangle$, and the
conductance is then given by the Landauer formula as $g={\rm
Tr}\,t^{\dag}t$. In particular, the point conductance between two edges
is ${\rm Tr}\,G(e_{\rm out},e_{\rm in};1)^{\dag}
G(e_{\rm out},e_{\rm in};1)$. 

The above defines the quantum problem which we wish to study. Our main
theorems will relate it to a classical problem, defined as follows. 
For each node in $\cal G$, adopt an arbitrary but fixed labelling of the
incoming edges $j\in\{1,\ldots,N\}$ and outgoing edges 
$i\in\{1,\ldots,N\}$, and denote the elements of the corresponding
$S$-matrix by $S_{ij}$. Note that $\det S=\pm1$, with the sign being
dependent on the choice of labelling.

Define a \em trail \em $\tau$ on $\cal G$ as a sequence of distinct edges
$(e_1,\ldots,e_{|\tau|})$ such that $e_k$ and $e_{k+1}$ are incoming and
outgoing edges of the same node, for each $1\leq k\leq |\tau|-1$. It is
a (rooted) closed trail if $e_{|\tau|}$ and $e_1$ also share the same node. 
Note that a trail cannot pass along a given edge more than once, but it
may pass through a given node any number of times, up to its order.

For a particular trail $\tau$, and a
particular node $n$, denote the 
set of incoming edges on $\tau$ by $J_{n;\tau}$, and the set of
outgoing edges by $I_{n;\tau}$. 
These are both \em ordered \em subsets of $\{1,2,\ldots,N\}$.
A given trail associates an element of
$I_{n;\tau}$ with each element of $J_{n;\tau}$, and vice versa, and thus may
be associated with a permutation $\pi_{n;\tau}$ of the ordered elements of
$J_{n;\tau}$. Denote the signature of this by $(-1)^{\pi_{n;\tau}}$. 
Let $\det S_{I,J}$ denote the minor
of $S$ restricted to the ordered subsets $I$ and $J$ of the outgoing and
incoming channels respectively. 

We are now ready to state

\noindent{\bf Theorem 1.} {\it The mean of $G(e_1,e_2;z)$ vanishes if
$e_1\not=e_2$, while in the case of equality it is given by
\begin{equation}
{\rm Tr}\,\overline{G(e,e;z)}=\left\{
\begin{array}{l@{\quad:\quad}l}
2-\sum_{\tau(e)}w_{\tau(e)}\,z^{2|\tau(e)|}&|z|<1\\
\sum_{\tau(e)}w_{\tau(e)}\,z^{-2|\tau(e)|}&|z|>1
\end{array}\right.
\end{equation}
where the sums are over all closed trails $\tau(e)$ rooted at $e$
and $w_{\tau(e)}$ is the weight of each,
given by the product over all the nodes on $\tau(e)$
of factors
\begin{equation}
\label{weight}
\Omega(I_{n;\tau};J_{n;\tau})\equiv
(-1)^{\pi_{n;\tau}}\,
\prod_{j\in J_{n;\tau}}S_{\pi_{n;\tau}(j),j}
\quad(\det S_{I_{n;\tau},J_{n;\tau}})
\end{equation}}

Note that the first two factors are the term in the expansion
of $\det S_{I_{n;\tau},J_{n;\tau}}$ corresponding to the permutation
$\pi_{n;\tau}$: if we were to sum (\ref{weight}) over all permutations, we
would obtain $\big(\det S_{I_{n;\tau},J_{n;\tau}}\big)^2$.
(\ref{weight})
is a generalisation of the main result of \cite{BCC}, which applies to
the case $N=2$. In this case, the elements of $S$ may be taken to be
$S_{11}=S_{22}=\cos\theta$ and $S_{12}=-S_{21}=\sin\theta$. If $J=\{1\}$
and $I=\{1\}$, or if $J=\{2\}$ and $I=\{2\}$, the weight is $\cos^2\theta$. 
If $J=\{1\}$ and $I=\{2\}$, or if $J=\{2\}$ and $I=\{1\}$,
it is $\sin^2\theta$. But if $J=\{1,2\}$ and $I=\{1,2\}$ or $\{2,1\}$, 
it is unity.

The next theorem gives the equivalent result for conductance
measurements. 

\noindent{\bf Theorem 2.} {\it The mean point conductance $\bar g$ between two
edges $e_{\rm in}$ and $e_{\rm out}$ is given by twice the sum over
all open trails on $\cal G$ connecting the two edges, 
each such path being weighted as for the closed loops in Theorem 1.}

Thus, if the weights on the trails
can be interpreted as a probability measure, the
mean conductance between two contacts 
is just twice the expected number of open trails which connect
them.

The weight for a single trail $\tau$ is given 
by a product of weights corresponding to each node thorough which $\tau$
passes, once the whole of $\tau$ is given. Alternatively, we may 
build up these weights as a product of factors incurred each time $\tau$
passes through a given node. For example, the first time it passes
through, entering via edge $j_1$ and leaving by edge $i_1$ it incurs a
weight, according to (\ref{weight}), 
of $\Omega(i_1;j_1)=S_{i_1,j_1}^2$.
If it passes though the same node again, this time entering along $j_2$
and leaving along $i_2$, it then incurs a conditional weight\footnote{This
assumes $S_{i_1,j_1}\not=0$. If the conditional weight can be
interpreted as a probability, $w(i_1,i_2;j_1,j_2)\leq1$ and therefore
it has a finite limit as $S_{i_1,j_1}\to0$. Even if this is not the case,
the unconditional weight $w(i_1,i_2;j_1,j_2)\,w(i_1;j_1)$ vanishes in
this limit.}
$w(i_1,i_2;j_1,j_2)=\Omega(i_1,i_2;j_1,j_2)/\Omega(i_1;j_1)$
and so on. In general
\begin{equation}
\label{cond}
w(i_1\ldots,i_p;j_1,\ldots,j_p)={\Omega(i_1\ldots,i_{p-1},i_p;j_1,\ldots,
j_{p-1},j_p)\over\Omega(i_1\ldots,i_{p-1};j_1,\ldots,j_{p-1})}
\end{equation}
The next theorem states that these conditional
weights are properly
normalised, in the sense that they give unity when summed over all
possible outcomes: 

\noindent{\bf Theorem 3.} {\it The weights $w(i_1\ldots,i_p;j_1,\ldots,j_p)$
satisfy
\begin{equation}
\sum_{i_p\not\in\{i_1,\ldots,i_{p-1}\}}w(i_1\ldots,i_p;j_1,\ldots,j_p)=1
\end{equation}}

Thus, as long as they are non-negative, they define a set of transition
probabilities for a discrete 
random process whereby the ensemble of trails may be
dynamically generated with the correct weights, the trail growing by one
unit at each time step. Since the weights at a given node depend on
whether (and how) it has been visited in the past, the process
may be thought of
as a history-dependent random walk. When all nodes have $N=2$, this is 
straightforward: the first time $\tau$ passes through a given node, it
incurs a factor $\cos^2\theta$ or $\sin^2\theta$. If it passes through a
second time, this factor is unmodified. 

\begin{figure}[h]
\centerline{\epsfig{width=8cm,file=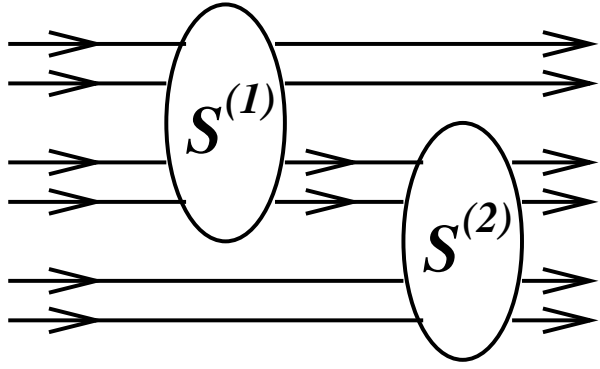}}
\caption{A reducible $S$-matrix.}
\label{reduce}
\end{figure}

We may ask whether this positivity can extend to nodes with $N\geq3$.
To answer this, we need a notion of reducibility. The $S$-matrix at 
a node with $N\geq3$ is said to be
\em reducible \em if it admits a factorisation of the form
$S=S^{(1)}S^{(2)}$, where (after a possible re-ordering of the incoming and
outgoing channel labels) the $N\times N$ $S^{(1)}$ and $S^{(2)}$ matrices
have the block diagonal forms
\begin{equation}
\label{red}
S^{(1)}=\left(\matrix{s^{(1)}_p&0\cr0&1_{N-p}\cr}\right)\qquad\quad
S^{(2)}=\left(\matrix{1_q&0\cr0&s^{(2)}_{N-q}\cr}\right)
\end{equation}
where $s^{(1)}_p$ and $s^{(2)}_{N-q}$ are orthogonal $p\times p$ and
$(N-q)\times(N-q)$ matrices respectively, and $p>q$. This is illustrated in
Fig.~\ref{reduce}. 
This procedure may be repeated. An $N\times N$
$S$-matrix is
said to be \em completely reducible \em if it can be factorised in this
way into $2\to2$ $S$-matrices.

\noindent{\bf Theorem 4.} {\it At a node with $N\geq3$, a sufficient
condition for the weights 
(\ref{weight}) to be all non-negative is that the $S$-matrix is
completely reducible. For $N=3$ and $4$ this is also necessary.}
Thus, in these cases, the network model on $\cal G$ could have been described  
on an equivalent graph with only $N=2$ nodes.

\section{Path integral representation.}

In the standard way, $G$ may be written as a path integral over
commuting (bosonic) variables.
The notation is a little complicated, but the basic idea is simple.
Label each end of a given directed edge $e$ by $e_R$ and $e_L$, 
in the direction of propagation $e_R\to e_L$. Introduce complex integration
variables ${b}_R(e)$ and ${b}_L(e)$, which are each 2-component
column vectors in SU(2) space, their components being labelled
${b}_{Ra}(e)$ and ${b}_{La}(e)$ respectively, where $a=1,2$. 
Then $G$ can be written as
\begin{equation}
\label{nd}
G(e_2,e_1;z)=\langle {b}_L(e_2){b}_L^{\dag}(e_1)\rangle
={\int\prod_e[d{b}_L(e)][d{b}_R(e)]{b}_L(e_2){b}_L^{\dag}(e_1)\,e^{W_b}
\over \int\prod_e[d{b}_L(e)][d{b}_R(e)]\,e^{W_b}}
\end{equation}
where 
$W_b=W_{\rm edge}+W_{\rm node}$ with
\begin{eqnarray}
W_{\rm edge}&=&z\sum_{e}{b}_L^{\dag}(e)U_e{b}_R(e)\\
W_{\rm node}&=&
\sum_{n}\sum_a\sum_{ij}{b}_{Ra}^*(e_i)
(S_n)_{ij}{b}_{La}(e_j)
\end{eqnarray}
and the integration is wrt the usual coherent state measure
\begin{equation}
\int[d{b}]=(1/\pi^2)\int e^{-{b}^{\dag}{b}}\prod_ad{\rm Re}\,{b}_a\,d{\rm
Im}\,{b}_a
\end{equation}
Note that there is a finite number of integrations, if $\cal G$ is
finite, and that no time-ordering necessary: we can imagine writing
everything out in terms of components, and all quantities in the path
integral are commuting. On a finite graph, only a finite number of
integrations is necessary. The exponentiation of $W_b$ correctly takes into
account the multiple traversing of edges by Feynman paths. 

The next step is to average over the quenched random variables $U_e$.
As usual, since these occur in both the numerator and denominator, this
is most easily done either by introducing replicas, or by adding 
an anticommuting (fermionic) copy
of the bosonic variables, making it supersymmetric. We opt for the
latter. 
Thus to each pair of complex integration variables $({b}^{\dag},{b})$
we introduce a pair of Grassmann variables $(\bar{f},{f})$
with corresponding labels, and we add to $W_b$ a term $W_f$ of identical
form with bosonic variables replaced by fermionic ones.
The Grassmann integration is defined 
by 
\begin{equation}
\int[d{f}]=\int d\bar{f} d{f} e^{-\bar{f}{f}}
\end{equation}
so that
\begin{eqnarray}
\int[d{f}]{f}=\int[d{f}]\bar{f}&=&0;\\
\int[d{f}]\,1=\int[d{f}]{f}\bar{f}&=&1
\end{eqnarray}

Integrating over the fermionic variables cancels the denominator in
(\ref{nd}), so that we may write
\begin{equation}
G(e_2,e_1)=
\int\prod_e[d{b}_L(e)][d{b}_R(e)][d{f}_L(e)][d{f}_R(e)]
{b}_L(e_2){b}_L^{\dag}(e_1)e^{W_b+W_f}
\end{equation}
However, $W_b+W_f$ is invariant under global supersymmetry, so $G$ may
equally well be expressed, for example,
as $\langle{f}_L(e_2)\overline{f}_L(e_1)\rangle$.

\subsection*{Quenched average.}
The average over the SU(2) matrix $U$
on a given edge has the form
\begin{equation}
\int dU\, \exp(z{b}^{\dag}_LU{b}_R+z\bar{f}_LU{f}_R)
\end{equation}
where the integral is with respect to
the invariant measure on SU(2), normalised
so that $\int dU=1$.  

\noindent {\bf Lemma 1:} \em The above integral equals
$1+\frac12z^2\det{\bf M}$, where $\bf M$ is the $2\times2$
matrix with components 
$M_{ij}={b}^*_{Li}{b}_{Rj}+\bar{f}_{Li}{f}_{Rj}$. \em

\noindent{\it Proof:}
Because the fermionic variables have only two
components, and any such component squares to zero, the expansion in
the fermionic part terminates:
\begin{equation}
\int dU\,\exp(z{b}^{\dag}_LU{b}_R)\left(1+z\bar{f}_LU{f}_R+\ffrac12
z^2(\bar{f}_LU{f}_R)^2\right)
\end{equation}
The first term, the purely bosonic integral, is identically equal to
unity. This 
follows from the observation that the integral is invariant under 
the substitutions ${b}_R\to \lambda V_R{b}_R$, ${b}_L^{\dag}\to
\lambda^{-1}{b}_L^{\dag}V_L^{\dag}$, where $V_L$ and $V_R$ are independent 
SU(2) matrices, and $\lambda$ is a complex number,
and there is no combination of ${b}_L^{\dag}$ and
${b}_R$ which has this property, save a constant.
However, an explicit proof is given in Appendix A.

The third, purely fermionic, term is also easy:
\begin{eqnarray}
(\bar{f}_LU{f}_R)^2&=&
(\bar{f}_{L1}U_{11}{f}_{R1}+\bar{f}_{L2}U_{21}{f}_{R1}+
\bar{f}_{L1}U_{12}{f}_{R2}+\bar{f}_{L2}U_{22}{f}_{R2})^2\\
&=&2(\bar{f}_{L1}U_{11}{f}_{R1})(\bar{f}_{L2}U_{22}{f}_{R2})
+2(\bar{f}_{L2}U_{21}{f}_{R1})(\bar{f}_{L1}U_{12}{f}_{R2})\\
&=&2\bar{f}_{L1}\bar{f}_{L2}{f}_{R2}{f}_{R1}(U_{11}U_{22}-
U_{12}U_{21})\\
&=&2\bar{f}_{L1}\bar{f}_{L2}{f}_{R2}{f}_{R1}\\
&=&\det\left(\matrix{\bar{f}_{L1}{f}_{R1}&\bar{f}_{L1}{f}_{R2}\cr
                     \bar{f}_{L2}{f}_{R1}&\bar{f}_{L2}{f}_{R2}\cr}
\right)
\end{eqnarray}
where the fourth line follows because $\det U=1$. The expression is
therefore independent of $U$, and the integration is then the same
as in the purely bosonic term, which gives a factor 1 as before.

The second term can also be worked out explicitly, but it easier to
invoke the supersymmetry, and simply add 
${b}^*_{Li}{b}_{Rj}$ to each element $\bar{f}_{Li}{f}_{Rj}$
of the above matrix. Note that the purely bosonic part of the determinant
vanishes. \qed

The result of the quenched average over the
SU(2) matrix on a given edge is therefore
\begin{equation}
\label{prop}
1+
\ffrac12z^2({b}^*_{L1}\bar{f}_{L2}-{b}^*_{L2}\bar{f}_{L1})
({b}_{R1}{f}_{R2}-{b}_{R2}{f}_{R1})
+z^2(\bar{f}_{L1}\bar{f}_{L2})({f}_{R2}{f}_{R1})
\end{equation}
The interpretation of this is clear: after averaging over the 
SU(2) matrices, the only paths which contribute are those in which
on each edge the allowed propagation is
of either the identity, a pair of fermions $f_1f_2$, or a boson-fermion pair
$(1/\sqrt2)(b_1f_2-f_1b_2)$. Note
that in each case the combinations in parentheses above are 
SU(2) singlets. 
Note also that, having averaged over the edge variables $U_e$, the
distinction between $L$ and $R$ is now immaterial, and we can henceforth
drop these labels.

The above result has several important
consequences. First, there is now only a finite number $3^{\cal N}$
of possibilities for propagation along the $\cal N$ edges of 
a finite graph $\cal G$. (This is equivalent to the result
of Gruzberg et al\cite{GLR} that the Hilbert space of the transfer matrix is
finite-dimensional.) 
Second, it is clear why the assumption that the scattering at the nodes
in diagonal in the SU(2) indices was not crucial: only the singlet
invariant amplitude matters.
Third, the only non-zero two-point functions 
with $e_2\not=e_1$ are
\begin{eqnarray}
\label{2pt}
&&\ffrac12\langle\left(b_{1}(e_2)f_{2}(e_2)-b_{2}(e_2)f_{1}(e_2)\right)
\left(b^*_{1}(e_1)\overline f_{2}(e_1)-b^*_{2}(e_1)
\overline f_{1}(e_1)\right)\rangle\\
&&=\langle f_{1}(e_2)f_{2}(e_2)\,\overline f_{2}(e_1)
\overline f_{1}(e_1)\rangle\\
&&=\overline{G_{11}G_{22}-G_{12}G_{21}}
=\overline{\det G(e_2,e_1;z)}
\end{eqnarray}

Let us for the moment take $z$ real. Then $G(e_2,e_1;z)$, as a sum over
Feynman paths, is a linear combination of SU(2) matrices with real
coefficients. Any such $2\times2$ matrix is itself proportional to an
SU(2) matrix, up to a real scalar (see Appendix.) Thus we may write
$G=\lambda\widetilde G$ where $\lambda$ is real and
$\widetilde G\in$ SU(2). Hence $\det G=\lambda^2$, and 
$G^{\dag}G=\lambda^2I$, so that ${\rm Tr}\,G^{\dag}G=2\det G$. The right hand
side is a polynomial in $z$. For general complex $z$ we have, therefore,
\begin{equation}
2\,\overline{\det G(e_2,e_1;z)}={\rm Tr}\,\overline{G(e_2,e_1;z^*)^{\dag}
G(e_2,e_1,z)}
\end{equation}
When $z=1$ this is the mean point conductance, which is therefore given,
up to a factor 2,
by the two-point functions in (\ref{2pt}). 

Since only SU$(2)$ singlets now propagate, it follows that two-point
functions like $\langle f_{a}(e_2)\overline f_{a}(e_1)\rangle
=\overline{G(e_2,e_1)}$ vanish if $e_2\not=e_1$. This is because, once
the matrices $U_e$ have been traced out, the supersymmetric path
integral possesses a \em local \em SU(2) gauge invariance under
$(b(e),f(e))\to (V_eb(e),V_ef(e))$ with $V_e\in$ SU(2). 

However, this does not apply if $e_2=e_1$. In fact, because of
(\ref{2pt}), it follows that
\begin{equation}
\overline{G(e,e;z)_{11}}=\langle f_{1}(e)\overline f_{1}(e)\rangle
=\langle f_{1}(e)f_{2}(e)\overline f_{2}(e)\overline f_{1}(e)\rangle
=\overline{\det G(e,e;z)}
\end{equation}

\section{Propagation through the nodes.}

In the last section, we showed that, for a graph $\cal G$ with $\cal N$
edges, the quenched average of the path integral can be written as a
sum of $3^{\cal N}$ terms, according to which of the three terms in 
(\ref{prop})
(corresponding to the propagation of a $bf$ pair, an $ff$ pair, or
the identity) is chosen on each edge. Let us now consider just one of these
terms, and one particular node.  

The contribution to the path integral from this node has the 
form
\begin{equation}
\label{ASA}
\prod_iA_{\alpha_i}(r_i)\,{\cal S}\,\prod_j\overline A_{\alpha_j}(r_j)
\end{equation}
where $A_1=1$, $A_2={f}_1{f}_2$ and
$A_3=(1/\sqrt2)({b}_1{f}_2-{b}_2{f}_1)$ 
and 
\begin{equation}
{\cal S}=\exp(\sum_{a=1}^2\sum_{i,j}({b}^*_{ia}S_{ij}{b}_{ja}
+\bar{f}_{ia}S_{ij}{f}_{ja}))
\end{equation}

In doing this, we have brought together in the path integral all the
factors associated with the given node. There is a subtlety,
however, because the boson-fermion variables $A_3$ and $\overline
A_3$ anticommute with each other. At a given node, we may arrange these
factors in the standard order determined by the fixed (but arbitrary)
labelling of the incoming and outgoing edges. For a given term out of
the $3^{\cal N}$ possibilities, this will introduce an overall factor $\pm1$.

Define a \em decomposition \em of the node as
a pairing of each outgoing edges $i$ with a unique incoming edge $j$.
This defines a permutation $\pi$ of the edge labels, whereby the outgoing
edge $i$ paired with the incoming edge $j$ is $\pi(j)$. 
Carried through for every node in turn, this decomposes $\cal G$ into a union
of disjoint directed
closed loops (and open paths if $\cal G$ is open), such that
every edge lies on just one loop or open path, and
each loop or open path may pass along a given edge no more than once.
The following Proposition shows that we are allowed to do
this inside the path integral, as long as we weight each decomposition 
correctly:

\noindent {\bf Proposition 1.} \em The result of performing the integration
over the variables $(b_j,f_j)$ and $(b_i^*,\overline f_i)$ in (\ref{ASA})
is the same as if $\cal S$ were replaced by 
\begin{equation}
\label{sameas}
\det S\,\sum_\pi(-1)^\pi\,\prod_{ij}\delta_{i,\pi(j)}S_{ij}
\delta_{\alpha_i,\alpha_j}
\end{equation}
that is, it is given by a weighted sum over all decompositions $\pi$. 
In each decomposition, each state
on the incoming edge $j$ propagates freely to $\pi(j)$. \em

\noindent{\it Proof:} 
Since the numbers of each component of both bosons and
fermions are the same in the incoming and outgoing channels, and bosons
are always paired with fermions, it follows that the numbers 
of $ff$ and $fb$ pairs are individually conserved at every node. 
Let us call the subsets of the $N$ outgoing channels occupied by an $ff$
pair, a $bf$ pair, or empty,
$FF$, $FB$, and $E$, respectively, and similarly for the incoming
channels, $\overline{FF}$, $\overline{FB}$ and $\overline E$. 
The integrations may now be performed, expanding $\cal S$ to second
order in the $S_{ij}$ and using Wick's theorem. Each fermion (boson) in
outgoing channel $i$, when contracted with a fermion (boson) in the
incoming channel $j$, gives (up to a sign) a factor $\delta_{ab}S_{ij}$.
The bosons in $FB$ may only contract onto the bosons in $\overline{FB}$,
but the complication is that some of the fermions in $FF$ may
contract onto fermions in $\overline{FB}$, and some of those in $FB$
may contract onto those in $\overline{FF}$. However, 
every set of possible contractions will involve each outgoing channel
in $FF\cup FB$ and each incoming channel in
$\overline{FF}\cup\overline{FB}$ exactly twice. Thus, if $\sigma$ denotes a
permutation of the channels in $FF\cup FB$, then, the general form of
the result will be
\begin{equation}
\sum_{\sigma,\sigma'}a_{\sigma,\sigma'}
\prod_{i\in FF\cup FB}S_{i,\sigma(i)}
\prod_{i'\in FF\cup FB}S_{i',\sigma'(i')}
\end{equation}
where the $a_{\sigma,\sigma'}$ are numerical coefficients. 

We have already introduced the notation $\det S_{I,J}$ for the
minor of $S$ restricted to the ordered 
subsets $I$ and $J$ of
outgoing and incoming channels. Now define $\perm S_{I,J}$ to be the
corresponding permanent,
that is, with all the terms having the same
sign $+1$. 
Then the claim is that the result of the integration is 
\begin{equation} 
\label{dDd}
\det S_{FF,\overline{FF}}\cdot\perm S_{FB,\overline{FB}}\cdot
\det S_{FF\cup FB,\overline{FF}\cup\overline{FB}}
\end{equation}
This expression has the correct properties in that:
(a) each channel index appears
exactly twice in each term; (b) $S_{ij}$ with $i\in FF$ and 
$j\in \overline{FB}$ (and also with $i\in FB$ and $j\in\overline{FF}$
occurs at most once; (c) it is symmetric under permutations of the channels
in $FF$, and separately in $\overline{FF}$; (d) it is antisymmetric under 
permutations of the channels in
$FB$, and separately in $\overline{FB}$; and (e) it
has the correct overall numerical coefficient. 

In order to prove (\ref{dDd}), it is helpful first to consider what happens if
each SU(2) singlet $\frac1{\sqrt2}(b_1f_2-b_2f_1)$ is replaced by 
$b_1f_2$ (and similarly for the conjugate variables in the incoming
channels.) In that case, the result follows immediately. The 
$f_{1i}$ with $i\in FF$ can contract only onto the $\overline f_{1j}$
with $j\in\overline{FF}$, giving the first factor in (\ref{dDd}). 
Similarly, the $b_{1i}$ with $i\in FB$ can contract only onto the
$b^*_{1j}$ with $j\in\overline{FB}$, giving the second factor. Finally
the $f_{2i}$ are free to contract onto any of the $\overline f_{2j}$, 
leading to the last factor. The reason that this result continues to
hold when each $b_1f_2$ is replaced by the singlet combination is the
local gauge invariance already alluded to: 
we could imagine multiplying the whole amplitude
by an independent SU(2) matrix in each channel, and averaging over
this. The final result, being gauge invariant, would not change, but it
would project $b_1f_2$ onto the singlet combination. 

We now need the following  property of O$(N)$ matrices:

\noindent {\bf Lemma 2.} \em If $S\in\mbox{O}(N)$, and $\det S'$ and $\det S''$ are
complementary minors of $S$, then $\det S'=\det S''\cdot\det S$. \em

\noindent{\it Proof:} This relies on the fact that if $S'$ has rank $p$,
and $T_{j_1\ldots j_p}$ is a tensor of rank $p$, then
$\epsilon_{j_i\ldots j_pj_{p+1}\ldots j_N}T_{j_1\ldots j_p}$ 
(where $\epsilon_{\ldots}$ is the Levi-Civita symbol)
transforms under proper rotations
as a tensor of rank $N-p$, and changes sign under parity. \qed

In our case this implies that 
$\det S_{FF\cup FB,\overline{FF}\cup\overline{FB}}=
\det S_{E,\overline E}\cdot\det S$, so that (\ref{dDd}) reads
\begin{equation}
\label{dDdd}
\det S_{FF,\overline{FF}}\cdot\perm S_{FB,\overline{FB}}\cdot
\det S_{E,\overline E}\cdot\det S
\end{equation}
Now look at (\ref{sameas}), inserted into the path integral instead of
$\cal S$. The Kronecker 
deltas which conserve the labels $\alpha$ restrict the
sum over permutations $\pi$ to those which map $\overline{FF}$ onto some
permutation $\pi_{FF}$ of
$FF$, $\overline{FB}$ onto some permutation $\pi_{FB}$ of $FB$, and 
so on. The signature $(-1)^\pi$ decomposes into a product of the
signatures of the three permutations. 
Now, since the $ff$ pairs propagate freely (and they commute among
themselves), the integrations over these variables give unity. 
The sum over $\pi_{FF}$ is therefore 
$\sum_{\pi_{FF}}(-1)^{\pi_{FF}}\prod_{i\in FF}S_{i,\pi_{FF}(i)}
=\det S_{FF,\overline{FF}}$, which gives the first factor in
(\ref{dDdd}).
Although the $bf$ pairs also propagate freely, they are fermionic, which
mens that their contractions give rise to an extra factor
$(-1)^{\pi_{FB}}$. On summing over all $\pi_{FB}$, we get the second
factor in (\ref{dDdd}). The remaining factors of $S_{\pi(j),j}$, with 
$j\in\overline E$, when summed over $\pi_E$, give the last factor.

We have shown the equivalence of the two expressions $\cal S$ and
(\ref{sameas}) at each node, for each of the $3^{\cal N}$ terms in the expansion
of the path integral. We may now restore the anticommuting $bf$ factors to their
original ordering in the path integral, thus removing the possible
overall sign. This concludes the proof of Prop.~1. \qed

\subsection*{Proof of Theorems 1 and 2.}

First consider the case when $\cal G$ is closed.
In Sec.~III it was shown that 
$\overline{G(e,e)}$ is given by 
the correlation function $\langle f_{L1}(e)f_{L2}(e)
\bar f_{L2}(e)\bar f_{L1}(e)\rangle$ in the supersymmetric path integral. 
By the results of the previous Section, this is given by a sum of terms
in which each edge except $e$ is occupied by either an
$ff$ pair, a $bf$ pair, or the identity (and $e$ 
is occupied only by an $ff$ pair.)
Moreover, the path integral is given by a sum of terms, each
corresponding to a decomposition of $\cal G$ into closed loops. Along
all but one of the closed loops can freely propagate an
$ff$ pair, giving an overall factor +1, an $bf$ pair, giving -1, or
the identity, giving +1. The first two contributions cancel, leaving a factor
+1 for each of these closed loops.
The exception is the unique loop which contains the edge $e$, which can
be thought of as a closed trail $\tau(e)$, rooted at $e$.
Along this only an $ff$ pair is allowed to propagate. 

Now sum over all decompositions of $\cal G$ which contain the specified
trail $\tau(e)$. At a given node $n$, $\tau(e)$ occupies the incoming edges
$J_{n;\tau}$ and the outgoing edges $I_{n;\tau}$. 
The sum in (\ref{sameas}) includes
only those permutations $\pi$ for which $\pi(J_{n;\tau})$ is some
permutation of $I_{n;\tau}$. This implies that 
$\pi$ acting on the complementary subset $\overline J_{n;\tau}$
is some permutation $\bar\pi$ of the complement $\overline I_{n;\tau}$.
If we now sum the factors of $S_{ij}$ in
(\ref{sameas}) with $i\in \overline I_{n;\tau}$ and 
$j\in\overline J_{n;\tau}$ 
over the permutations $\bar\pi$, weighted by $(-1)^{\bar\pi}$, we get
$\det S_{\overline I_{n;\tau},\overline J_{n;\tau}}$. Using Lemma 2
again, this equals $\det S_{I_{n;\tau},J_{n;\tau}}\cdot\det S$. The latter
factor of $\det S$ combines with explicit one in (\ref{sameas}) to give
unity. The remaining factors then give the weight (\ref{weight})
of the node $n$ on the trail $\tau(e)$. This proves Theorem 1. \qed 

Theorem 2 follows similarly. For an open graph, 
${\rm Tr}\,\overline{G^{\dag}(e_2,e_1)G(e_2,e_1)}$ is given by a sum of
decompositions of $\cal G$ as before, into closed loops as well as open
paths which connect the incoming and outgoing external edges.
Along these propagate either $ff$ pairs, $bf$ pairs, or the identity, with
weights at each node given by (\ref{sameas}). In each decomposition,
there is a unique open trail $\tau$ from $e_1$ to $e_2$, carrying an
$ff$ pair. The other open paths must carry the identity, otherwise the path
integration over the free bosonic and fermionic variables at their ends
would give zero. They therefore contribute a factor 1. All the other
closed loops also contribute a factor 1 after the cancellation between
the $ff$ and $bf$ pairs which propagate around each of them. We are left
with a single $ff$ pair propagating along $\tau$. The summation over all
the decompositions of $\cal G$ containing a given open trail $\tau$ then
gives a factor $\det S_{I_{n;\tau},J_{n;\tau}}\cdot\det S$ at each node
as above. This proves Theorem 2. \qed

\section{Probabilistic interpretation.}
\subsection*{Normalisation.}
We now prove Theorem 3, which states
that the weights $\Omega(I,J)$ in Theorem 1 lead, if non-negative, 
through (\ref{cond}) to correctly normalised transition
probabilities $w(i_1,\ldots,i_p;j_1,\dots,j_p)$
for the trail $\tau(e)$ interpreted as a
classical random walk on the edges of $\cal G$. 
A necessary and sufficient condition for this is 
\begin{equation}
\label{NSC}
\sum_{i_p\notin\{i_1,\ldots,i_{p-1}\}}\Omega(i_1,\ldots,i_{p-1},i_p;
j_i,\ldots,j_{p-1},j_p)=
\Omega(i_1,\ldots,i_{p-1};j_1,\ldots,j_{p-1})
\end{equation}
Without loss of generality, we may relabel the rows and columns of $S$
so that $i_k=k$ for $1\leq k\leq p-1$, and $j_k=k$ for $1\leq k\leq p$.
Notice that we can remove the restriction on the sum over $i_p$ because
the summand formally vanishes whenever $1\leq s_p\leq p-1$.
The index $i_p$ occurs on the left hand side of (\ref{NSC}) in the
factor $S_{i_p,p}$ as well as in each term of the expansion of the
minor $\det S_{\{1,\ldots,i_p\};\{1,\ldots,p\}}$, where it occurs as
$S_{i_p,k}$ with $1\leq k\leq p$. Thus the sum over $i_p$, in each term
in the expansion of the determinant, has the form
$\sum_{i_p}S_{i_p,p}S_{i_p,k}=\delta_{pk}$, from the orthonormality
of the rows of $S$. The coefficient of this term is just the subminor
$\det S_{\{1,\ldots,p-1\};\{1,\ldots,p-1\}}$ which occurs on the
right hand side of (\ref{NSC}). All the remaining factors 
$\prod_{1\leq k\leq p-1}S_{k,k}$ are the same on both sides. This
demonstrates the validity of (\ref{NSC}) and thus Theorem 3. \qed

\subsection*{Positivity of the weights.}
Although we have argued that the weights $\Omega$ appearing in
Theorem 1 are normalised, they may only be interpreted as probabilities
if they are all non-negative. This places strong constraints on the
$S$-matrix at each node. 

Taking first the case when the sets $I$ and $J$ comprise all the
outgoing and incoming edges of the node, we see that the weights are all
non-negative if and only if every term in the expansion of $\det S$ has
the same sign, or vanishes. In fact, this is also a sufficient condition
for all the weights to be non-negative when $I$ and $J$ are proper
subsets. This is because, by Lemma 2, $\det S_{I,J}$ is, up to a factor
$\det S=\pm1$, the same as its conjugate minor, and therefore each term
in (\ref{weight}) is, up to an overall sign, a sum of a subset of 
terms in the
expansion of $\det S$. They therefore all have the same sign, or vanish,
if this is true of the individual terms in the expansion. 

For $N=2$, this is always the case. If $\det S=1$, we can write
$S=\left(\matrix{\cos\theta&\sin\theta\cr-\sin\theta&\cos\theta}\right)$,
so that the terms in the expansion are $(\cos^2\theta,\sin^2\theta)$;
or if $\det S=-1$ we can write
$S=\left(\matrix{\sin\theta&\cos\theta\cr\cos\theta&-\sin\theta}\right)$,
in which case they are $(-\sin^2\theta,-\cos^2\theta)$. 
However, for an orthogonal matrix with $N>2$, this constraint becomes
nontrivial. 
\subsubsection{$N=3$.}
Consider first the case $N=3$. 
It is elementary to show that if the $3!$ terms in the expansion of
the determinant of \em any \em $3\times3$ matrix all have the sign (or
vanish) then there must be at least one vanishing element. 
For consider the \em product \em of all these terms.
This contains each element $S_{ij}$ exactly
twice. There are six terms in all, and three of these, corresponding to the
odd permutations, occur with minus signs. Hence the product of all the
terms is $-\prod_{i=1}^3\prod_{j=1}^3S^2_{ij}\leq0$. 
This would be impossible if all the $S_{ij}$ were non-vanishing.

\begin{figure}[h]
\centerline{\epsfig{width=8cm,file=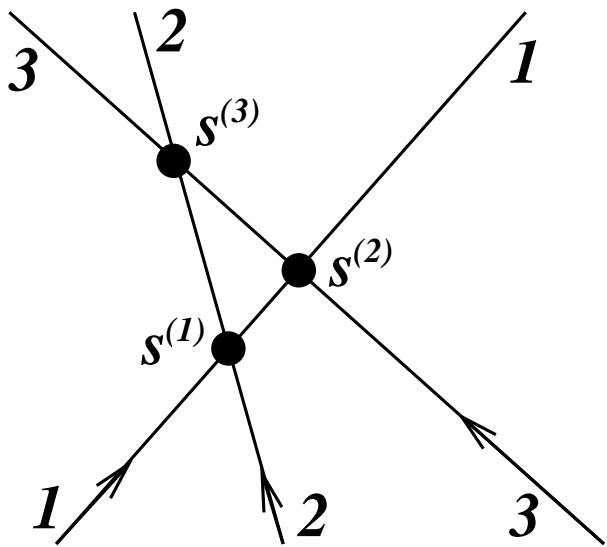}}
\caption{Diagram illustrating Euler angle representation of an O$(3)$ 
$S$-matrix.}
\label{euler}
\end{figure}

Now any O(3) rotation can be composed of three suitable O(2) rotations about
different axes, for example through the Euler angles. 
This composition may, in general, be pictured using a diagram like
that in Fig.~\ref{euler}. 
Each intersection of lines labelled by $i$ and $j$
corresponds to an O(2) rotation in the $ij$ plane, represented by an
O(2) matrix $s^{(a)}$ with $a=1,2,3$. The element
$S_{ij}$ of the full O(3) matrix is given by a sum over directed paths
from $j$ to $i$ in the diagram, each path being weighted by a product
of the appropriate O(2) matrix elements. For example,
\begin{eqnarray}
S_{13}&=&s^{(2)}_{13}\label{simple}\\
S_{31}&=&s^{(3)}_{32}s^{(1)}_{21}+s^{(3)}_{33}s^{(2)}_{31}s^{(1)}_{11}
\end{eqnarray}
Each topologically distinct way of drawing and labelling Fig.~\ref{euler}
corresponds
to a different but equivalent Euler angle parametrisation.

We can always draw the diagram so that the matrix
element which vanishes by the above argument
(in this example $S_{13}$) is given by a simple
form like (\ref{simple}). This implies that $s^{(2)}_{13}=s^{(2)}_{31}=0$, and 
therefore that $s^{(2)}_{11}=s^{(2)}_{33}=1$ (note that $s^{(2)}$ can
always be chosen as a proper rotation.) This means that we can picture
the lines 1 and 3 simply crossing at the vertex (2), and that the full
O(3) rotation reduces into a product of just two O(2) rotations, as in
the definition (\ref{red}) of reducibility.
\subsubsection{$N>3$.}
\begin{figure}[h]
\centerline{\epsfig{width=8cm,file=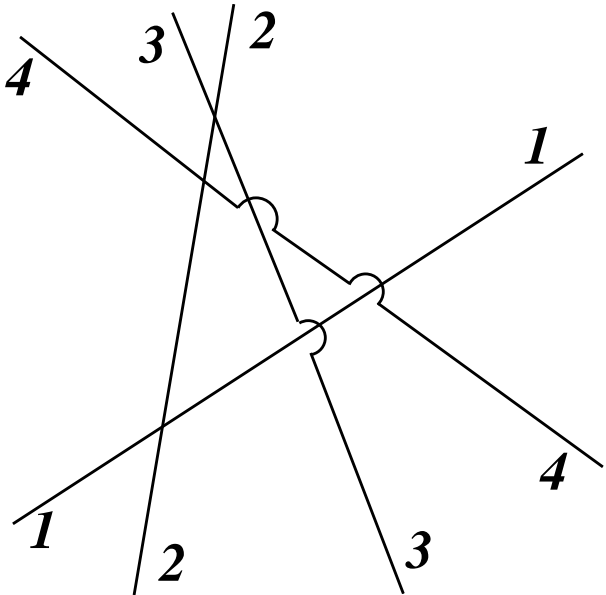}}
\caption{Euler angle representation for $N=4$: an example where
vanishing elements force the matrix to be completely reducible.
In this case $S_{14}=S_{13}=S_{24}=0$.}
\label{euler4}
\end{figure}

An O$(N)$ matrix has several distinct but equivalent
Euler angle representations as
a composition of $\frac12N(N-1)$ O$(2)$ rotations, which may be pictured
using a generalisation of Fig.\ref{euler}. An example for $N=4$ is shown in
Fig.~\ref{euler4}. 
In such a diagram any given line intersects each of
the others exactly once. If the matrix is completely reducible
there is at least one representation which has
a tree structure, that is, contains no cycles. An example is shown in
Fig.~\ref{euler4}. In this case, many elements of $S$ must vanish, and those
which do not are each given by a single term which is a product of
O$(2)$ matrix elements along a single possible path through the diagram.

Complete reducibility is a sufficient condition for the weights in
(\ref{weight}) all to be non-negative. One way to see this is to note that
we can in this case decompose the node into a tree of $2\to2$ nodes.
The internal edges of this tree can be made to carry an arbitrary
SU$(2)$ matrix, which can however always be set equal to 1 by making a
gauge transformation on the SU$(2)$ matrices on the
incoming and outgoing edges of the node
(this is not always possible if there are cycles.) We may therefore
introduce such matrices on each internal 
edge of the tree and integrate over them
without changing the problem. Thus the weights for the node are products
of weights for $2\to2$ nodes, which we have already argued are always
non-negative. 

Next we consider whether this condition is necessary. Consider the terms
in the expansion of $\det S$ which contain a factor $S_{11}S_{22}\ldots
S_{N-3,N-3}$. The coefficient of this term is the $3\times3$
minor $\det S_{I,J}$ with $I=J=\{N-2,N-1,N\}$. 
The above Lemma about $3\times3$ matrices then shows that 
either this submatrix has at least one vanishing element, or the 
product $S_{11}S_{22}\ldots S_{N-3,N-3}$ vanishes. In general, every 
$3\times3$ submatrix of $S$ must have at least one vanishing element, or
every term in the expansion of its complementary minor must vanish.

For $N=4$, this implies that there must
be at least 3 vanishing elements, not all in the same row or column.
By considering the different cases, together with a suitably chosen
Euler angle representation, it is possible to show that in each case a
sufficient number of the O$(2)$ matrix elements must vanish that the
diagram breaks up into a tree. This shows that, for $N=4$, the condition
of complete reducibility is also necessary for non-negative weights.
However, we have not found a general argument for all $N$ and indeed
there may be exceptions. What can be shown straightforwardly is that $S$
must have at least $N-1$ vanishing elements. 
\vspace{1cm}

\noindent{\it Acknowledgements.} 
The author would like to thank John Chalker, Ilya Gruzberg and Martin
Zirnbauer for helpful comments and criticism.
This work was supported in part by the EPSRC under Grant GR/R83712/01
The initial phase was carried out while the author
was a member of the Institute for Advanced Study.
He thanks the School of Mathematics and the School of Natural Sciences
for their hospitality.
This stay was supported by the Bell Fund, the James D.~Wolfensohn Fund,
and a grant in aid from the Funds for Natural Sciences.

\appendix
\section*{Some properties of SU$(2)$ matrices}
We show explicitly that the integral 
\begin{equation}
\label{dU}
{\cal I}\equiv\int\,dU\,\exp(zb_L^{\dag}Ub_R)=1
\end{equation}
Any SU$(2)$ matrix may be parametrised as
$U=\exp(i\alpha{\bf\sigma}\cdot{\bf n})
=\cos\alpha+i{\bf\sigma}\cdot{\bf n}\sin\alpha$. The Haar measure is
then
\begin{equation}
\int\,dU=(2\pi^2)^{-1}\int_0^\pi \sin^2\alpha\,d\alpha\,\int d\Omega_{\bf n}
\end{equation}
The exponent in (\ref{dU}) has the form $A\cos\alpha+i{\bf n}\cdot{\bf
B}$ where $A=zb_L^{\dag}b_R$ and ${\bf B}=zb_L^{\dag}{\bf\sigma}b_R$.
Note that ${\bf B}^2=A^2$. Although these are in general complex, since
$\cal I$ is an analytic function of each of their components, we can
first assume they are real. Then, without loss of generality, we can
assume that $\bf B$ is real and points in the $z$-direction. Then 
\begin{equation}
{\cal I}=(1/\pi)\int_0^\pi\sin^2\alpha d\alpha\int_{-1}^1d(\cos\theta)
\exp\big(A(\cos\alpha+i\cos\theta\sin\alpha)\big)
\end{equation}
The integral over $\cos\theta$ is simple, and the result may be expanded
in a power series in $A$. All terms, save that $O(A^0)$, then vanish on
integration over $\alpha$. 

We show that any real linear combination of SU$(2)$ matrices is itself,
up to a real constant, an SU$(2)$ matrix. From the above representation,
it may be written as
\begin{equation}
G=\sum_ja_j\cos\alpha_j+i\sum_ja_j{\bf n}_j\cdot{\bf\sigma}\sin\alpha_j
\end{equation}
which has the form $A+iB{\bf N}\cdot{\bf\sigma}$ where $A$ and $B$ are
real, and ${\bf N}$ is another unit 3-vector. Writing $A=\rho\cos\alpha$
and $B=\rho\sin\alpha$ then gives the required result.

\end{document}